# Spectra of Young Galaxies

*Stéphane Charlot*

Astronomy Department and Center for Particle Astrophysics,
University of California, Berkeley CA 94720, USA

## 1 Introduction

The expected widespread population of high-redshift analogs or progenitors of present-day disk galaxies has still not been observed in emission. This leaves large uncertainties on the appearance of young galaxies (see [Wh1] and [Pr1]). Early models predicted that "primeval galaxies" undergoing strong starbursts at redshifts $z \gtrsim 2$ should be detectable at magnitudes $R \sim 21-23$, but deep spectroscopic surveys to $B \lesssim 22.5$, $I \lesssim 22.1$, and $K \lesssim 20$ have not revealed such population of forming galaxies ([PP1], [Me1], [CEBTB1], [CGHSHW1], [LTHCL1]). Partridge and Peebles (see [PP1]) pointed out that Ly$\alpha$ emission could be the most prominent and easily detectable signature of primeval galaxies. The reason for this is that the ionizing radiation from young stars in galaxies should lead to a strong and narrow Ly$\alpha$ line by recombination of the hydrogen in the ambient interstellar medium, which would be more readily visible than continuum radiation against the sky noise. A few galaxy-like objects have been discovered at redshifts $z \lesssim 4$, which occasionally show strong Ly$\alpha$ emission ([Pr1], [Sp1], [MLGTS1], [Ret1]). However, most of these objects are peculiar, and their connection to present-day galaxies is not at all clear. In fact, all blank sky searches for Ly$\alpha$ emission from ordinary galactic disks at high redshifts have given null results, suggesting that young galaxies form stars slowly or in large volumes, or that Ly$\alpha$ photons are absorbed by dust (see [BW1], and references therein).

Independently, much has been learned on the distribution of HI in the universe at redshifts $z \lesssim 3.5$ from absorption-line studies of distant quasars (see for example [PWRCL1]). The strongest absorption lines are attributed to the damped Ly$\alpha$ systems, that are generally interpreted as the best candidates for ordinary galactic disks at high redshift ([LWTLMH1]). These have observed HI column densities $N_{\rm HI} \gtrsim 2 \times 10^{20}$ cm$^{-2}$, and their abundances in heavy elements and dust at $z \approx 2.5$ amount to about 10% of the values in the Milky Way ([PFB1], [PSHK1]). The damped Ly$\alpha$ systems do not show Ly$\alpha$ emission at the level expected from young, dust-poor disk galaxies (see [PSHK1]). Alternatively, the evolution with redshift of the gas density integrated over all HI absorbers, $\Omega_{\rm HI}(z)$, provides some constraint on the global depletion of cold gas though star formation in the universe since $z \approx 3.5$ ([LWT1], [PF1]). Absorption-line systems of distant quasars have also been used successfully as a way to select normal galaxies in emission out to $z \approx 1.6$ against the dominant population of relatively nearby blue galaxies that dominate galaxy number counts at faint magnitudes ([Be1], [SD1], [SDP1], [AESD1]). These studies have shown that field galaxies with luminosities around $L^*$ exhibit only little evolution in their space density, luminosity, and optical/infrared colors at redshifts $0.2 \lesssim z \lesssim 1.6$.

Recent observations therefore seem to indicate that the formation and evolution of normal disk galaxies has been less spectacular than originally thought. In what follows, we explain how the apparent lack of Ly$\alpha$ emission from young galaxies at high redshift is probably mainly a consequence of the relatively brief periods in which primeval galaxies are dust-free, and hence Ly$\alpha$-bright. In fact, most present observational constraints on young galaxies appear to be in agreement with the predictions of theories based on hierarchical clustering, in which galaxies form slowly and relatively recently. The very blue, primeval galaxy phase expected at the onset of star formation would then be faint and short-lived. Since at redshifts $z \gtrsim 2$ galaxies are expected to be hard to detect, one may think of using population synthesis models to trace back the early history of star formation from observations at lower redshifts. We also discuss below the limitations of this approach.

## 2 Ly$\alpha$ Emission from Young Galaxies

The observed Ly$\alpha$ emission from a young galaxy depends on the star formation rate and initial mass function (IMF), but also on several other factors: the contributions by supernova remnants and active galactic nulcei, the orientation of the galaxy, and absorption by dust. The contribution to the Ly$\alpha$ emission by stars can be estimated using stellar population synthesis models. We assume for the moment that circumstellar HII regions are the only sources of Ly$\alpha$ photons and that the column density of the ambient HI is large enough that case B recombination applies ($N_{\rm HI} \gtrsim 10^{17}\,{\rm cm}^{-2}$) but otherwise ignore the effects on the interstellar medium on the transfer of Ly$\alpha$ photons. Under these "minimal" assumptions and using recent population synthesis models, Charlot and Fall have shown that the Ly$\alpha$ emission from a galaxy depends sensitively on the age and IMF slope, even when the star formation rate is constant ([CF1]). The dependence on the IMF upper cutoff and metallicity, on the other hand, are much weaker. Thus, only a rough estimate of the Ly$\alpha$ equivalent width of a young, dust-free galaxy is permitted, about $50 - 120\,{\rm Å}$.

We now briefly review the other factors that can affect the observed Ly$\alpha$ emission from a young galaxy (see [CF1] for more details). Shull and Silk have computed the time-averaged, Ly$\alpha$ luminosity of a population of Type II supernova remnants using a radiative-shock code with low metallicity ([SS1]). Their results indicate that the contribution to the Ly$\alpha$ emission by supernova remnants is always less than the contribution by stars (typically 10% for a solar-neighborhood

IMF) and can therefore be neglected. Active Galactic Nuclei (AGNs) are another potential source of ionizing radiation in a galaxy. We assume for simplicity that the spectrum of an AGN can be approximated by a power law $f_\nu \propto \nu^{-\alpha}$ with an index blueward of Ly$\alpha$ in the range $1 \lesssim \alpha \lesssim 2$. If we also assume that the AGN is completely surrounded by HI, that case B recombination applies, and that absorption by dust is negligible, then the Ly$\alpha$ equivalent width is $827\alpha^{-1}(3/4)^\alpha$ Å, or 600 Å for $\alpha = 1$ and 200 Å for $\alpha = 2$. The fact that most bright quasars have observed Ly$\alpha$ equivalent widths in the range $50 - 150$ Å could reflect a partial covering of the AGNs by HI clouds in the broad-line regions (in fact, some ionizing radiation escapes from quasars), attenuation of the Ly$\alpha$ emission by dust, or orientation effects (see below). Thus, in principle, AGNs can produce higher Ly$\alpha$ equivalent widths than stellar populations. However, the presence of an AGN in a galaxy is usually revealed by other readily identifiable signatures: strong emission lines of highly ionized species (CIV, HeII, etc.) and broad emission lines with velocity widths several times larger than those expected from the virial motions within galaxies.

The Ly$\alpha$ photons produced in galaxies will suffer a large number of resonant scatterings in the ambient neutral atomic hydrogen. In the absence of dust, this would lead to no net enhancement of the angle-averaged Ly$\alpha$ emission from a galaxy or of the total Ly$\alpha$ emission from a sample of randomly-oriented galaxies. However, since the Ly$\alpha$ line is emitted more isotropically than the continuum, the Ly$\alpha$ equivalent width of an individual galaxy will decrease as it is viewed more nearly edge-on. For example, in the idealized case of a plane-parallel slab, the ratio of the observed to angle-averaged Ly$\alpha$ equivalent width will decrease from 2.3 to 0 for viewing angles to the normal ranging from 0° to 90°. The resonant scattering of Ly$\alpha$ photons by HI also increases enormously their chances of absorption by dust grains. The attenuation is expected to be important when the dimensionless dust-to-gas ratio, defined in terms of the extinction optical depth in the $B$ band by $k \equiv 10^{21}(\tau_B/N_{\rm HI})\,{\rm cm}^{-2}$, exceeds the critical value $k_{\rm crit} \approx 0.01(N_{\rm HI\perp}/10^{21}\,{\rm cm}^{-2})^{-4/3}(\sigma_V/10\,{\rm km\,s}^{-1})^{2/3}$ (see [CF1]). In this expression, $N_{\rm HI\perp}$ and $\sigma_V$ are the face-on column density and line-of-sight velocity dispersion of HI. For reference, the dust-to-gas ratio in the Milky Way and Large and Small Magellanic Clouds are, respectively, $k \approx 0.8$, $k \approx 0.2$, and $k \approx 0.02$, and the face-on HI column densities within the optically visible regions of most spiral galaxies lie in the range $10^{20} \lesssim N_{\rm HI\perp} \lesssim 10^{21}\,{\rm cm}^{-2}$. Thus, we expect $k \gtrsim k_{\rm crit}$ unless the dust-to-gas ratio is much smaller than the value in the Milky Way. In particular, some attenuation of the Ly$\alpha$ emission by dust is expected in the damped Ly$\alpha$ systems, since $N_{\rm HI} \gtrsim 2 \times 10^{20}\,{\rm cm}^{-2}$ and $k \approx 0.1$ (although there may be a large dispersion around this value; see [PFB1] and [PSHK1]). Moreover, the attenuation of Ly$\alpha$ emission by dust depends sensitively on the structure of the interstellar medium in a galaxy. In the case of a multiphase medium, the transfer of Ly$\alpha$ photons will depend largely on the topology of the interfaces between HI and HII regions (see [Sp2] and [Ne1]). As a result of these complications, it is nearly impossible to deduce star formation rates from the observed Ly$\alpha$ emission of a galaxy.

Charlot and Fall (see [CF1]) have used the above arguments to interpret the observations of and searches for Lyα emission from nearby star-forming galaxies, damped Lyα systems, blank sky, and the companions of quasars and damped Lyα systems. Their results indicate that, when Lyα emission is weak or absent, as is the case in most star-forming galaxies at low redshifts and in damped Lyα systems at high redshifts, the observed abundance of dust is sufficient to absorb most of the Lyα photons. On the other hand, when Lyα emission is strong, the presence of highly ionized species, large velocity widths, or nearby quasars indicate that much of the ionizing radiation may be supplied by AGNs. The hope has always been that the searches for Lyα emission at high redshifts would reveal a population of primeval galaxies, in which the abundances of heavy elements and hence dust were low enough that most of the Lyα photons could escape. Such a population may exist at some redshifts. However, since the Lyα emission is attenuated when the dust-to-gas ratio exceeds $1-10\%$ of the value in the Milky Way, a typical galaxy probably spends only the first few percent of its lifetime in a Lyα-bright phase. We therefore expect primeval galaxies, as defined above, to be relatively rare at most redshifts, consistent with the null results of all searches to date.

## 3 What Should Young Galaxies Look Like?

If galaxies undergoing their first episodes of star formation in the young universe resemble nearby examples of starburst galaxies, one would expect them to exhibit very blue continua and strong Balmer emission lines. However, the most distant counterparts of nearby disk galaxies found by association with quasar absorption-line systems do not show such extreme signatures ([SD1], [AESD1]). Instead, out to $z \approx 1.6$, normal field galaxies appear to display strikingly little evolution in their space density, luminosity, and optical/infrared colors. Other constraints on the history of star formation in galaxies, and hence on their appearance, may be obtained from the evolution of the integrated neutral gas density of the universe. Recent surveys indicate that the total amount of gas in damped Lyα systems at a redshift $z \approx 3.5$ could be nearly as high as the total density of luminous material in present-day galactic disks ([LWT1], [Wo1]). This would imply that most stars have formed relatively recently, consistent with the expectation from theories of galaxy formation based on hierarchical clustering (although a more precise interpretation of the observed $\Omega_{\rm HI}$ evolution requires including the effect of quasar obscuration by dust in damped Lyα systems; see [PF1]).

A natural question to ask, then, is what do models accommodating present observational constraints predict for the appearance of very young galaxies? To answer this question, we have computed the early spectral evolution of a spiral galaxy using a combination of the [KWG1] semi-analytic model of galaxy formation and the [BC1] population synthesis code. The star formation rate can be adjusted to reproduce both the evolution of $\Omega_{\rm HI}(z)$ at $z \lesssim 3.5$ in a standard cold dark matter universe and the spectral energy distribution of a typical nearby spiral galaxy at $z = 0$ (see [KC1]). The early spectral evolution of this model is presented in Figure 1. Figure 1a shows that the galaxy undergoes an extremely blue phase at $z > 3$, during which the spectrum resembles the observed spectrum of the nearby starburst galaxy Mrk 710 with strong H-Balmer and oxygen emission lines (from [VC1]). At this time, the young galaxy could qualify as a "primeval galaxy" as defined earlier. However, the phase is short-lived ($\lesssim 10^7$ yr) and very faint ($K > 25$). Then, the onset of evolved supergiant, asymptotic giant branch, and red giant branch stars reddens the spectrum substantially. At $z = 1.5$, twenty percent of the stars present at $z = 0$ have formed, and the model galaxy in Figure 1b has $K \approx 20$ and $R - K \approx 3.2$, i.e., a spectrum only moderately bluer than that of the nearby spiral galaxy NGC 6181 (from [Ke1]). These predicted colors, which ignore reddening by dust, are interestingly close to the observed $K \approx 19.5$ and $R - K \approx 4$ of galaxies discovered at $z \approx 1.5$ in association with quasar absorption-line systems ([SD1]). Hence, the present results would reinforce the suggestion that the spectra of normal disk galaxies have evolved only moderately for much of their lifetime. The extremely blue phase at the onset of star formation, which might coincide with a Lyα-bright phase, is expected to be much fainter and short-lived.

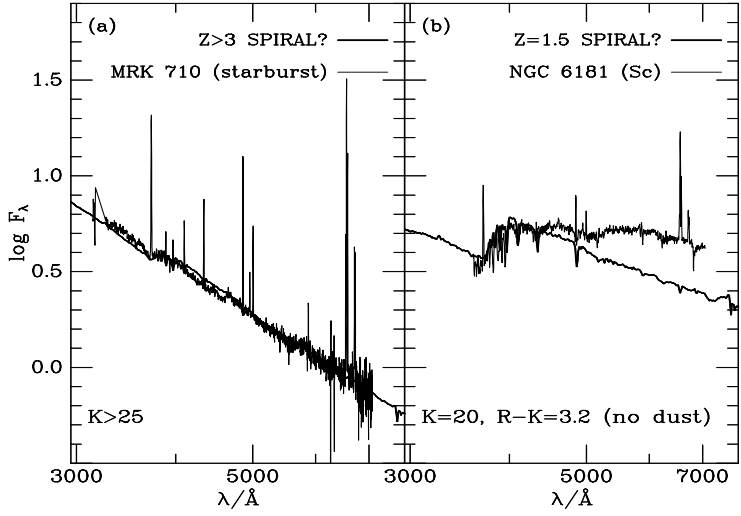

**Fig. 1.** (a) Possible appearance of a progenitor spiral galaxy at $z > 3$ satisfying the current observational constraints on the evolution of $\Omega_{\rm HI}(z)$ and the observed properties of the Milky Way in a standard cold dark matter universe (thick line) compared to the observed spectrum of the nearby starburst galaxy Mrk 710 (thin line). The model spectrum does not include emission lines (see text for more details). (b) Same model galaxy as in (a) viewed at $z = 1.5$ and compared to the observed spectrum of the nearby spiral galaxy NGC 6181. The spectra in (a) and (b) are in the rest frames of the galaxies, and the predicted apparent $K$ magnitudes (and $R - K$ color at $z = 1.5$) are indicated at the bottom.

## 4 Tracing Back the History of Star Formation in Galaxies

Since young galaxies may be hard to detect at redshifts beyond $z \sim 2$, an alternative is to try and trace back the earlier history of star formation from observations at lower redshifts. The conventional approach to this problem is to use stellar population synthesis models and search for the evolution of the star formation rate that will reproduce the observed spectral characteristics of galaxies. We now exemplify the difficulties in this approach using the recent population synthesis models of [BC1]. Figure 2a shows the spectral evolution of a burst stellar population computed for a Salpeter IMF with lower and upper cutoffs 0.1 $M_\odot$ and 100 $M_\odot$, respectively. At $10^6$ yr, the spectrum is entirely dominated by short-lived, young massive stars on the main sequence. Then, the ultraviolet light declines and the spectrum reddens rapidly as lower-mass stars progressively complete their evolution. The most remarkable feature in Figure 2a is the nearly unevolving shape of the optical to near-infrared spectrum at ages from 4 to 19 Gyr. This "age degeneracy" hampers the dating of passively evolving stellar populations such as elliptical galaxies from the single knowledge of the continuum spectrum, or equivalently, of broad-band colors (see [CS1], and references therein). The high luminosity of young massive stars also complicates the determination of the past history of star formation in spiral galaxies. This is illustrated in Figure 2b, in which we show the spectral decomposition of a present-day spiral galaxy represented by a model with constant star formation seen at 13.5 Gyr. The ultraviolet to visible spectrum is strongly dominated by massive, main-sequence and supergiant stars. Furthermore, as seen above, the contribution by evolved asymptotic giant branch and red giant branch stars is age-degenerated. Therefore, from the continuum spectrum of a spiral galaxy, one can at best obtain the ratio of the current to past-averaged star formation rate (see for example [KTG1]).

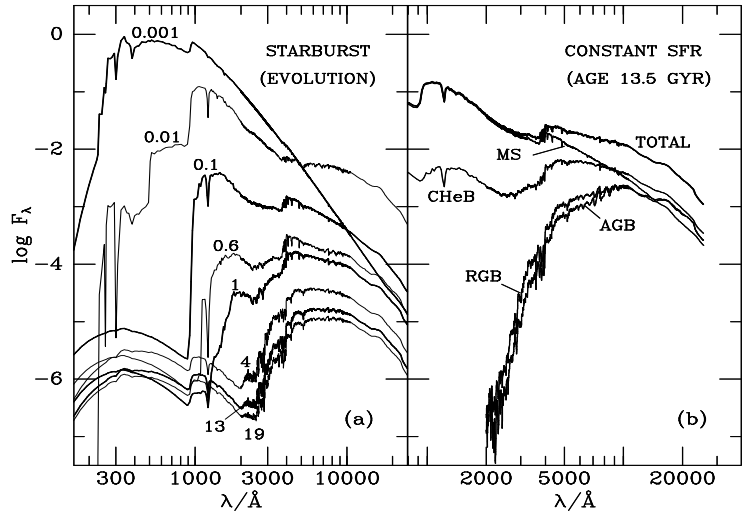

**Fig. 2.** (a) Spectral evolution of an instantaneous-burst stellar population computed with the population synthesis models of [BC1]. The stellar population has a Salpeter IMF with lower and upper cutoffs 0.1 $M_\odot$ and 100 $M_\odot$, respectively. Ages are indicated next to the spectra (in Gyrs). (b) Spectral decomposition of a present-day spiral galaxy represented by a model with constant star formation rate for 13.5 Gyr. The total spectrum is shown, together with the contributions main sequence (MS), supergiant (CHeB), asymptotic giant branch (AGB), red giant branch (RGB) stars. The IMF is the same as in (a).

Another important limitation in determining the past history of star formation in galaxies is the influence of metallicity. At fixed mass, stars of lower metallicity are hotter and evolve more rapidly than stars of higher metallicity. Thus, the age assigned to a stellar population on the basis of its observed colors is a function of metallicity. Table 1 illustrates the dependence on metallicity of the turnoff age (and mass of the turnoff star) at fixed turnoff temperature for a burst stellar population (after [SSMM1]). The assigned age can vary by a factor of up to five when the metallicity changes from 100% to 5% of solar. This is the "age-metallicity" degeneracy. In reality, the numbers in Table 1 are only roughly indicative of the age-metallicity degeneracy in interpreting galaxy continuum spectra because the colors of stars at fixed temperature also change slightly with metallicity. The most recent models of stellar population synthesis now include the effect of metallicity variations on the spectral evolution of galaxies (e.g., [Wo2], [BCF1]). These models can illustrate more accurately the age-metallicity degeneracy of spectral fits of galaxies.

**Table 1.** Dependence on metallicity of the turnoff age (and mass of the turnoff star) at fixed turnoff temperature for a burst stellar population (after [SSMM1]).

| $\log(T_{\rm TO}/{\rm K})$ | $Z = Z_\odot$ | $Z = 0.05 Z_\odot$ |
|---|---|---|
| 4.0 | 0.4 Gyr (3.0 $M_\odot$) | 1.3 Gyr (1.7 $M_\odot$) |
| 3.8 | 2.7 Gyr (1.5 $M_\odot$) | 15. Gyr (0.8 $M_\odot$) |

Some additional information on the past history of star formation in galaxies may be learned from the stellar absorption lines of hydrogen and of other prominent atoms and molecule such as Mg, $Mg_2$, Fe, Ca, Na, Sr, and CN. For ex-ample, main-sequence A and B stars (that have lifetimes $\lesssim 2$ Gyr) are expected to strengthen the H-Balmer series and weaken the prominent metallic lines in the integrated galaxy spectrum. This can be most simply illustrated by considering the case of intermediate-age stellar populations in early-type galaxies. There is growing photometric and dynamical evidence that many E/S0 galaxies have formed stars only a few billion years ago ([Pi1], [Ro1], [SS2], and references therein). The prototypical example is the dwarf elliptical galaxy M 32, which is believed to have undergone substantial star formation until only about 5 Gyr ago (e.g., [Oc1]). Figure 3a illustrates how spectral absorption features such as the Balmer H$\delta$ equivalent width can be used to detect late bursts of star formation in early-type galaxies with colors otherwise typical of old, passively evolving stellar populations. The solid lines correspond to a model elliptical galaxy formed in a major burst at age $t = 0$, on which a new burst involving 10% of the final mass is added at an age of 6 Gyr. At ages $t > 6.5$ Gyr, the $U - B$ and $V - K$ colors of this model differ by less than 0.1 mag from the values in the absence a second burst (shown by the dashed lines). However, the H$\delta$ equivalent width continues to evolve significantly for nearly 1 Gyr because of the presence of A and B stars. Pickles and Rose have shown, using other stellar absorption lines, how similar diagnostics can be used to discriminate between different generations of stars in E/S0 galaxies and, to some extent, to untangle the competing effects of age and metallicity on the spectra (see [Pi1] and [Ro1]). Worthey has also recently produced a comprehensive study of the absorption-line characteristics of old stellar populations and their dependence on age and metallicity ([Wo2], see also [WFG1]).

We now briefly exemplify some implications of these arguments for the history of star formation in early-type galaxies (see [CS1] for more details). In the upper panel of Figure 3b we have compiled estimates of the typical fraction of optical light accounted for by intermediate-age stars in normal E/S0 galaxies at redshifts $z \lesssim 0.4$ from several studies of stellar absorption-lines strengths ([Ro1], [Pi2], [CS2]). At $z \gtrsim 0.1$, most galaxies were selected in clusters (the estimates are based on all red galaxies with luminosities $\lesssim L^*$ for which spectra were available at each redshift). In each case, the range of intermediate ages attributed to the stars is indicated in billion years. In the lower two panels of Figure 3b, we have reexpressed these constraints on the ages of stellar populations into constraints on the redshifts of formation for two cosmologies using the [BC1] population synthesis models. A flat universe with $q_0 = 0.5$ and $h = 0.45$, and an open universe with $q_0 = 0.1$ and $h = 0.55$ (where $h = H_0/100\,\mathrm{km\,s^{-1} Mpc^{-1}}$). Both correspond to a present age of the universe of about 15 Gyr and lead to similar predictions: the mass fraction of stars formed in E/S0 galaxies has decreased smoothly with time, from about 8% at $z \approx 1$ to less than 1% at $z \approx 0$ (see also [Pi2] and [SS2]). This evolution of the star formation rate inferred from stellar absorption-line studies is in reality a mean evolution averaged over large redshift intervals, as the horizontal error bars indicate. It does not imply that E/S0 galaxies should form stars at all times. In fact, the ages of intermediate-age stars estimated from absorption-line strengths in the spectra of galaxies at low redshifts are uncertain by a few billion years. As Figure 3a shows, after a galaxy undergoes a burst of star formation, the colors reach the values characteristic of old, passively evolving stellar populations in less than 1 Gyr. Thus, although galaxies at low redshift may present similar signatures of past star formation, there should be a dispersion in the ages and hence colors of the progenitor galaxies at high redshift around the value corresponding to the mean epoch of star formation estimated in the lower panels of Figure 3b. The relevance of this result for the evolution of galaxies in clusters has been investigated by [CS1] (see also [BBRT1]).

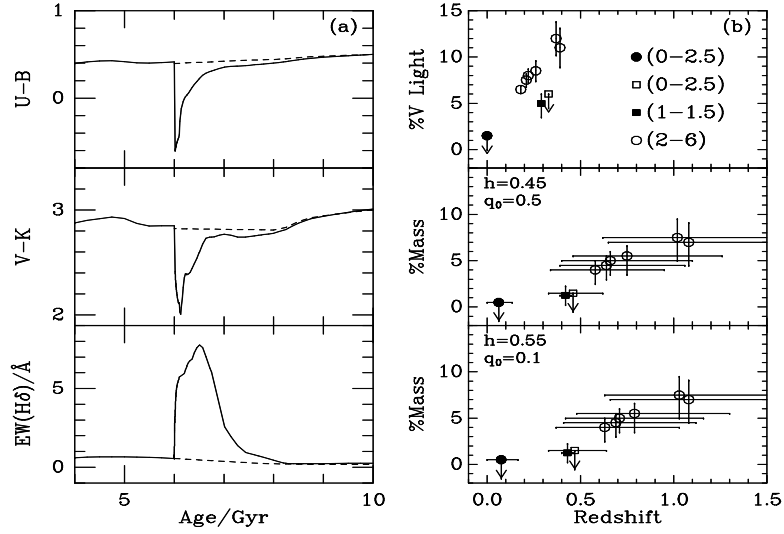

**Fig. 3.** (a) Evolution of the $U-B$ and $V-K$ colors and H$\delta$ absorption equivalent width of a model elliptical galaxy formed in a single burst at $t=0$ (dashed line) and of a similar model galaxy on which a new burst involving 10% of the final mass is added at an age of 6 Gyr (solid line). (b) Upper panel: observed contributions by intermediate-age stars to the $V$ luminosity of E/S0 galaxies at $z \approx 0$ and in low-redshift clusters inferred from stellar absorption-line studies (see text for sources). Different symbols correspond to different studies or different ranges of intermediate ages (indicated in Gyrs). Lower panels: mass fraction of stars formed in the progenitors of E/S0 galaxies as a function of redshift derived for two cosmologies from the observations shown in the upper panel. The horizontal error bars follow from the uncertainties on the ages of stars detected in low-redshifts E/S0 galaxies. The vertical error bars follow from the uncertainties on the determination of the contribution by these stars to the mass for the allowed range of ages.

Unfortunately, stellar absorption-line strengths cannot yet be used to trace back the history of star formation in spiral galaxies. The reason for this is that the best-known stellar absorption features arise at ultraviolet and optical wave-lengths, where the spectral signatures of old and intermediate-age stars in spiral galaxies are hidden by the strong continuum light of young massive stars (see Fig. 2b). The situation may soon be improved, as substantial progress is underway to understand the infrared spectral signatures of old stars in star-forming galaxies ([LR1]). However, tracing back the onset of star formation in normal disk galaxies appears to be a long way ahead.